
\hfill NUB 3054/92
$$     $$
\centerline{\bf Compact Lattice QED and the Coulomb Potential}
$$     $$
\centerline{Y.N. Srivastava, A. Widom and M.H. Friedman}
\centerline{Physics Department, Northeastern University}
\centerline{Boston, Massachusetts, U.S.A.}
\centerline{and}
\centerline{Physics Department \& INFN., University of Perugia}
\centerline{Perugia, Italy}
\centerline{and}
\centerline{O. Panella}
\centerline{Centre de Physique Theorique, CNRS Luminy}
\centerline{Marseille, France}
$$     $$
\centerline{\it ABSTRACT}
$$     $$

The potential energy of a static charge distribution on a lattice is
rigorously computed in the standard compact quantum electrodynamic
model. The method used follows closely that of Weyl for ordinary
quantum electrodynamics in continuous space-time. The potential energy
of the static charge distribution is independent of temperature and
can be calculated from the lattice version of Poisson's equation. It
is the usual Coulomb potential.

\vfill \eject

The notion of approximating gauge field theories in continuous
space-time by discrete lattice gauge models grew (in part) with
the desire to compute numerically the consequences of quantum
gauge theories$^{(1-5)}$. This might be accomplished without recourse to
perturbative methods. However, some features of the discrete lattice
version of QED theory appear more difficult than in the continuous
space-time theory. Such appears to be the case for the force law
between charges in compact lattice quantum electrodynamics$^{(6-13)}$.

For quantum electrodynamics in continuous space-time, the law of
force between two static charges is that of Coulomb, while the law
of force between two steady state current distributions follows from
Ampere's law. For lattice compact quantum electrodynamics, the Ampere
law can become somewhat difficult to compute due to the inherent
nonlinear form of the field equations. On the other hand, the lattice
compact quantum electrodynamic analog of the Coulomb law can be
directly computed. This lattice Coulomb law calculation is the purpose
of the work which follows.

The compact quantum electrodynamic lattice model (to be discussed) is
here described in the canonical Hamiltonian form. The cubic lattice
of the theory (three space dimensions) has vertices located
at positions $\{{\bf x}\}$. The components of a vector field at a vertex
live on the links $\{L\}$ connecting neighboring vertices. In addition
to the links connecting neighboring vertices, one must consider
the plaquettes $\{P\}$. These are the square areas bounded by four
links. An alternative view employs the dual lattice with
vertices located at $\{{\bf x^*}\}$ which are the center points of
the cubes on the original lattice. The links on the dual
lattice are in one to one correspondence to the plaquettes
on the original lattice. Each link on the dual lattice punctures a unique
plaquette area on the original lattice.

The electric field components live on the links of the original lattice.
The magetic field components live on the links of the dual lattice, or
equivalently on the plaquettes of the original lattice as the Faraday
law magnetic flux through the plaquette area.

Perhaps the most simple manner to describe QED on a lattice is to
employ a ``quantum circuit array'' viewpoint. (We use units
where $\hbar=1$ and $c=1$.) The (temporal gauge) relation between
the electric field and the vector potential,
$$
{\bf E}({\bf x})=-\dot{{\bf A}}({\bf x}), \eqno(1)
$$
on the lattice appears as an expression for the voltage across a link
$$
V_L=-\dot{\phi }_L . \eqno(2).
$$
The energy stored in the electric field is defined by assigning (to
each link) a capcitance C, so that the energy becomes
$$
K=(1/2)C\sum_L V_L^2 . \eqno(3)
$$

If we sum the voltages around the links (which form the boundary of a
plaquette), then the lattice version of Faraday's law is that we must
obtain the rate of change of the magnetic flux through the plaquette.
Formally, Faraday's law on the lattice arises as a consequence of the
definition of magetic flux $\Phi_P$ through a plaquette. It is
$$
exp(iq\Phi_P)=\prod_{L\in \partial P} exp(iq\eta_L\phi_L), \eqno(4)
$$
where $q$ is the ``charge'' unit of the theory, $\partial P$ is the
boundary of the plaquette, and the coefficients $\eta_L =\pm 1$ dictate
the orientaion of the plaqutte area. That we are discussing compact
QED on the lattice now makes an appearance since the inductance
$\Lambda $ (assigned to each plaquette) represents a nonlinear circuit
element; i.e. the magnetic field energy is given by
$$
U=(1/q^2 \Lambda )\sum_P [1-cos(q\Phi_P)]. \eqno(5)
$$

In total, the Lagrangian ($L=K-U$) of the model follows from
Eqs.(2),(3) and (5) to be
$$
L=(1/2)C\sum_L \dot{\phi}_L^2 -
(1/q^2\Lambda )\sum_P [1-cos(q\Phi_P)], \eqno(6)
$$
together with Eq.(4). The implied Hamiltonian is
$$
H=-(1/2C)\sum_L (\partial /\partial \phi_L)^2+
(1/q^2\Lambda )\sum_P [1-cos(q\Phi_P)]. \eqno(7)
$$

The wave functions of the theory have coordinates that live on a circle
and are thereby magnetic flux quantum ($2\pi /q$) periodic
$$
\psi(...,\phi_L+(2\pi /q),...)=\psi(...,\phi_L,...). \eqno(8)
$$
The inner product
$$
<\psi_f,\psi_i>=\oint\psi_f^*(...,\phi_L,...)\psi_i(...,\phi_L,...)
\prod_L(q d\phi_L/2\pi ), \eqno(9)
$$
completes the definition (quite rigorously for a finite lattice) of the
Hilbert space of states.

In the infinite lattice limit, there is one vertex per cube, three
links per cube and three plaquettes per cube. On the other hand, there
are only two linearly independent magnetic flux plaquette variables per
cube. This arises because the magnetic field is {\it transverse}
in the lattice sense to be described more fully below.
Furthermore, if one starts at time zero in a subspace of states
spanned by wave functions which depend only on the magetic flux
coordinates
$$
\psi_o =\Psi(...,\Phi_P, ...) \eqno(10)
$$
then one remains in this subspace at all future times. This is a
consequence of the Hamiltonian Eqs.(4) and (7). This subspace
corresponds to having {\it no charges} present on the vertices. The
magnetic flux coordinates here completely determine the dynamics.
Let us consider this in more detail.

With link unit vectors defined as
$$
{\bf n}_1=(1,0,0),\ {\bf n}_2=(0,1,0),\ {\bf n}_3=(0,0,1), \eqno(11a)
$$
each link on the lattice is determined by a vertex and a unit vector
$$
L\mapsto ({\bf x},{\bf n}),\ \ V_L=a{\bf n\cdot E}({\bf x})
=(i/C)(\partial /\partial \phi_L), \eqno(11b)
$$
where $a$ is the lattice spacing. The lattice divergence of the electric
field at a vertex ${\bf x}$ is defined as
$$
div{\bf E}({\bf x})=(1/a)\sum_{i=1}^3
{\bf n}_i {\bf \cdot }[{\bf E}({\bf x})-
{\bf E}({\bf x}-a{\bf n}_i)]. \eqno(12)
$$
On the dual lattice we have the identities
$$
exp(iq\Phi_P )=exp[iqa^2{\bf n\cdot B}({\bf x}^*)], \eqno(13a)
$$
$$
div^*{\bf B}({\bf x}^*)=0. \eqno(13b)
$$
The subspace of states spanned by functions of the magnetic field only,
as in Eq.(10), are precisely those which obey the Gauss law with no
charges at vertices
$$
div{\bf E}({\bf x})\psi_o=0. \eqno(14)
$$

Now suppose that we place charges on a subset of vertices
$\{{\bf x}_j\}$. The charge density (in terms of the discrete
$\delta $ function) is given by
$$
\rho ({\bf x})=(q/a^3)\sum_j z_j\delta({\bf x},{\bf x}_j), \eqno(15a)
$$
$$
z_j=\pm 1,\pm 2,\pm 3,\pm 4,... \ . \eqno(15b)
$$
To compute the wave functions with the external charges we must solve
the Gauss law equation
$$
div{\bf E}({\bf x})\psi =\rho ({\bf x})\psi . \eqno(16)
$$
To do so, we adopt the method used by Weyl$^{(18)}$ in continous space-time
QED to the problem at hand, {\it i.e.}, compact lattice QED$^{(19)}$. We first
decompose  the voltage accross a link into a {\it transverse} and {\it
longitudinal}  part;
$$
V_L=-\dot{\bar\phi }({\bf x},{\bf n})+
\dot{\chi }({\bf x})-\dot{\chi }({\bf x}+a{\bf n}), \eqno(17a)
$$
$$
exp[iq\bar{\phi}]\equiv
exp[iq\bar{\phi }({\bf x},{\bf n})]=
exp[iaq{\bf n\cdot \bar{A}}({\bf x})],\eqno(17b)
$$
$$
div{\bf \bar A}({\bf x})=0, \eqno(17c)
$$
$$
div{\bf E}({\bf x})
=(i/a^3)[\partial /\partial \chi ({\bf x})]. \eqno(17d)
$$
Only the transverse link coordinates contribute to the Faraday law
magnetic fux through a plaquette. Thus, Eq.(4) may be written in
terms of transverse link coordinates as
$$
exp(iq\Phi_P)=\prod_{L\in \partial P}exp(iq\bar{\phi }_L). \eqno(18)
$$

{}From Eqs.(17) and (6) we may write the Lagrangian as a ``radiation''
plus an ``Coulomb'' part
$$
L=(1/2)C\sum_L\dot{\bar\phi}_L^2-
(1/2q^2\Lambda )\sum_P[1-cos(q\Phi_P)]+L_{Coulomb}, \eqno(19a)
$$
$$
L_{Coulomb}=(1/2)C\sum_{\bf x}\sum_{i=1}^3
[\dot{\chi }({\bf x})-\dot{\chi }({\bf x}+a{\bf n}_i)]^2. \eqno(19b)
$$
The Hamiltonian has a similar decomposition
$$
H=H_{radiation}+H_{Coulomb}, \eqno(20a)
$$
$$
H_{radiation}=-(1/2C)\sum_L (\partial /\partial \bar{\phi}_L)^2+
(1/q^2\Lambda )\sum_P [1-cos(q\Phi_P)], \eqno(20b)
$$
$$
H_{Coulomb}=-(1/2C)\sum_{\bf xy}G({\bf x}-{\bf y})
[\partial^2/\partial \chi({\bf x})\partial \chi({\bf y})], \eqno(20c)
$$
where $G({\bf x}-{\bf y})$ is the lattice Coulomb Green's function to be
discussed in more detail below. Here we note that the subspace of wave
functions of the form
$$
\psi =exp[iq\sum_j z_j\chi ({\bf x}_j)]\Psi (...,\Phi_P,...), \eqno(21)
$$
solve the Gauss law Eq.(16) in virtue of Eqs.(10),(14),(15) and (17d).

While the radiation problem in compact lattice QED is quite complicated,
$$
H_{radiation}\Psi (...,\Phi_P,...)=E\Psi (...,\Phi_P,...), \eqno(22)
$$
radiation is decoupled from the Coulomb energy (for static charges if
not for dynamical Fermions$^{(14-17)}$).
The Coulomb energy is simply
$$
\Delta E_{Coulomb}=(q^2/2C)\sum_{ij}z_iz_jG({\bf x}_i-{\bf x}_j). \eqno(23)
$$
In going from the Lagrangian Eq.(19b) to the Hamiltonian Eq.(20c) one
must compute the lattice Coulomb Greens function
$$
-\nabla^2G({\bf x}-{\bf y})=\delta({\bf x},{\bf y}), \eqno(24a)
$$
in which the {\it lattice} Laplace operator is here defined as
$$
-\nabla^2f({\bf x})=\sum_{i=1}^3[2f({\bf x})-
f({\bf x}+a{\bf n}_i)-f({\bf x}-a{\bf n}_i)]. \eqno(24b)
$$
Eq.(24a) may be solved by Fourier transformation as
$$
G({\bf x})=(1/2)\oint [\prod_{i=1}^3 (d\theta_i/2\pi)]
e^{i({\bf \theta \cdot x}/a)}/[3-\sum_{j=1}^3cos(\theta_j)] . \eqno(25)
$$

It is proved in the mathematical literature on lattice random walks
that
$$
|{\bf x}|G({\bf x})\rightarrow (a/4\pi)\ \
(|{\bf x}|\rightarrow \infty ). \eqno(26)
$$
Our {\it genral proof} of the Coulomb law for compact lattice
QED is thus completed. The physical motivation of the proof (which Weyl
used in ordinary space-time QED) is simply as follows: If one enforces
the Gauss law on the physical wave functions, then the Coulomb law will
follow. That simple reasoning is sometimes obscured in the path integral
formalism unless sufficient attention is paid to the lattice Helmholtz
theorem.

We make no statements as to nature of the free energy for the nonlinear
radiation  (transverse) degrees of freedom in lattice compact QED. But
the Coulomb law for the longitudinal field components is on solid
ground for all temperatures. In Fig.1, we compare the continuous
space-time Coulomb  potential to the Coulomb law obtained on an
``infinite'' lattice as in Eq.(25). The convergence to the continuous
space-time Coulomb law takes about four lattice lengths. On the other
hand, suppose that the calculation of $G({\bf x},{\bf y})$ were performed
using a finite lattice with $N\times N\times N$ lattice sites.
At least $N\sim1,000$ would be required to obtain adequate numerical
evidence$^{(20)}$ for a long range Coulomb potential.

\vfill \eject
\centerline{\bf FIGURE CAPTION}
\vskip 0.4 in
\par \noindent
Fig.1: The lattice Coulomb potential in the ${\bf n}_1=(1,0,0)$ direction
is compared with that obtained in the continuous space-time form of the
Coulomb potential energy $(q^2z_1z_2/4\pi r)$.

\vskip 0.7 in
\centerline{\bf REFERENCES}
\vskip 0.4 in
\par
\item{[1]}K.~G.~Wilson, {\it Phys. Rev.} {\bf D10}, 2445 (1974).
\item{[2]}L.~P.~Kadanoff, {\it Rev. Mod. Phys.} {\bf 49}, 267 (1977).
\item{[3]}M.~Creutz, {\it Quarks, Gluons and Lattices},
 Cambridge University Press, (1983).
\item{[4]}J.~Kogut and L.~Susskind, {\it Phys. Rev.}, {\bf D9} 3501 (1974).
\item{[5]}J.~Frohlich and T.~Spencer, {\it Commun. Math. Phys.} {\bf 83},
411 (1982) .
\item{[6]}M.~Creutz, L.~Jacobs and C.~Rebbi, {\it Phys. Rev.} {\bf D20},
1915 (1979) .
\item{[7]}A.~H.~Guth, {\it Phys. Rev.} {\bf D15}, 2291 (1980) .
\item{[8]}T.~A.~DeGrand and D.~Toussaint, {\it Phys. Rev.} {\bf D22},
2478 (1980) .
\item{[9]}B.~Svetitsky, S.~D.~Drell, H.~R.~Quinn and M.~Weinstein, {\it
Phys. Rev.} {\bf D22}, 490 (1980).
\item{[10]}R.~Gupta, M.~A.~Novotny and R.~Cordery, {\it Phys. Lett.} {\bf
B172}, 86 (1986) .
\item{[11]}H.~Kleinert and W.~Miller, {\it Phys. Rev. Lett.} {\bf 56},
11 (1986).
\item{[12]}A.~N.~Burkitt, {\it Nuclear Physics}, {\bf B270[FS16]}, 575
(1986).
\item {[13]}C.~B.~Lang, {\it Nuclear Physics} {\bf B280[FS18]}, 255 (1987)
\item {[14]}J.~B.~Kogut and E.~Dagotto, {\it Phys. Rev. Lett.}  {\bf 59}, 617
(1987).
\item{[15]}M.~Okawa, {\it Phys. Rev. Lett.} {\bf 62},1224 (1989) .
\item{[16]}V.~Azcoiti, G. Di Carlo and A.~F.~Grillo, {\it Phys. Rev. Lett.}
{\bf 65}, 2239 (1990) .
\item{[17]}S.~Hashimoto, M.~Kikugawa and T.~Muta, {\it Phys. Lett.} {\bf
B254}, 449 (1991) .
\item{[18]}H.~Weyl, {\it Group Theory and Quantum Mechanics}, Dover
Publications, New York (1950).
\item{[19]}A.~M.~Polyakov, {\it Gauge Fields and Strings}, Harwood Academic
Publishers, New York (1987)
\item{[20]}D.~E.~Knuth, {\it The Art of Computer Programming}, Vol. II
Addison Wesley Publications, New York. (1969).
cf. after Eq. (3.53), page 46.
\bye